\documentclass{article}
\usepackage{spconf,amsmath,graphicx}
\usepackage[colorlinks,linkcolor=blue]{hyperref}
\usepackage{mwe}

\usepackage{times}
\usepackage{epsfig}
\usepackage{graphicx}
\usepackage{amsmath}
\usepackage{amssymb}
\usepackage{algorithm}
\usepackage{algorithmic}
\usepackage{mathrsfs}
\usepackage{multirow}

\usepackage{amsthm}
\theoremstyle{plain}

\usepackage[section]{placeins}

\usepackage[]{animate}

\usepackage{graphicx}
\usepackage{animate}

\usepackage{float}

\usepackage{multirow}
\usepackage{rotating}
\usepackage{wrapfig}
\usepackage{lineno}
\usepackage{epsfig}
\usepackage{amsmath,amssymb}
\usepackage{color}
\usepackage{booktabs}       
\usepackage{amsfonts,amsmath}       
\usepackage{nicefrac}       
\usepackage{microtype} 
\usepackage{float}
\usepackage{amsmath}
 
\usepackage{enumerate}
 
\title{Successive Subspace Learning for Cardiac Disease Classification\\ with Two-phase Deformation Fields from Cine MRI}
\name{
	\parbox{\linewidth}{\centering Xiaofeng Liu$^{1}$, Fangxu Xing$^{1}$, Hanna K. Gaggin$^{2}$, C.-C. Jay Kuo$^{3}$, {Fellow, IEEE,} \\ Georges El Fakhri$^{1}$, {Fellow, IEEE,} Jonghye Woo$^{1}$, {Senior Member, IEEE}}
}

\address{$^1$Dept. of Radiology, Massachusetts General Hospital and Harvard Medical School, Boston, MA, USA\\
$^2$Dept. of Medicine, Massachusetts General Hospital and Harvard Medical School, Boston, MA, USA\\
$^3$Dept. of ECE, University of Southern California, Los Angeles, CA, USA}
 
\begin{document}
%
\maketitle

\begin{abstract} 
Cardiac cine magnetic resonance imaging (MRI) has been used to characterize cardiovascular diseases (CVD), often providing a noninvasive phenotyping tool.~While recently flourished deep learning based approaches using cine MRI yield accurate characterization results, the performance is often degraded by small training samples. In addition, many deep learning models are deemed a ``black box," for which models remain largely elusive in how models yield a prediction and how reliable they are. To alleviate this, this work proposes a lightweight successive subspace learning (SSL) framework for CVD classification, based on an interpretable feedforward design, in conjunction with a cardiac atlas. Specifically, our hierarchical SSL model is based on (i) neighborhood voxel expansion, (ii) unsupervised subspace approximation, (iii) supervised regression, and (iv) multi-level feature integration. In addition, using two-phase 3D deformation fields, including end-diastolic and end-systolic phases, derived between the atlas and individual subjects as input offers objective means of assessing CVD, even with small training samples. We evaluate our framework on the ACDC2017 database, comprising one healthy group and four disease groups. Compared with 3D CNN-based approaches, our framework achieves superior classification performance with 140$\times$ fewer parameters, which supports its potential value in clinical use.

\end{abstract} 
%

\section{Introduction} 

Assessment of cardiovascular disease (CVD) using cine magnetic resonance imaging (MRI) (e.g., multi-slice 2D cine MRI) has been used to visualize and quantify structure and function of the beating heart~\cite{liu2021segmentation}. While manual evaluation of CVD using these complex spatiotemporal data is time-consuming and often non-reproducible, automatic CVD assessment can efficiently guide treatment, which is significant in reducing the risk of sudden death and the severity of symptoms.


Prior work on early diagnosis and prognosis of CVD using cardiac cine MRI~\cite{ammar2021automatic,liu2021segmentation} depends on the accurate segmentation of the left ventricle in end-diastolic (ED) and end-systolic (ES) phases~\cite{gaggin2013biomarkers} to derive clinical parameters, such as ejection fraction (EF). However, EF or related volumetric measures often lack sensitivity and discrimination ability. For the past several years, deep learning (DL)-based approaches \cite{goodfellow2016deep} have been actively developed in this area, demonstrating superior performance in phenotyping tasks. For example, convolutional neural networks (CNN)-based approaches have been the main workhorse behind this endeavor~\cite{shen2017deep}. 

Although DL-based approaches have yielded outstanding performance for computer-aided diagnosis or prognosis, several difficulties arise, which hinders their wide adoption for clinical applications \cite{shen2017deep}. First, DL-based approaches usually rely on massive labeled training datasets~\cite{goodfellow2016deep}; collecting and accurately labeling a sufficiently large number of training datasets for clinical applications pose challenges due partly to costly labeling or privacy concerns over sensitive patient information~\cite{liu2022memory}. Given a small number of training data, CNN-based models with a myriad of parameters are likely to overfit the training data, thus leading to performance degradation in deployment, when using datasets with a different distribution, compared with training data. More importantly, many DL models are considered a ``black-box"~\cite{kuo2016understanding,wang2022assessment}, for which DL models remain largely elusive in how the models yield a prediction and how reliable they are. The aforementioned issues motivate us to develop an interpretable framework that is compatible with a limited number of training datasets for clinical implementations.

\begin{figure*}[t]
  \centering 
\includegraphics[width=18cm]{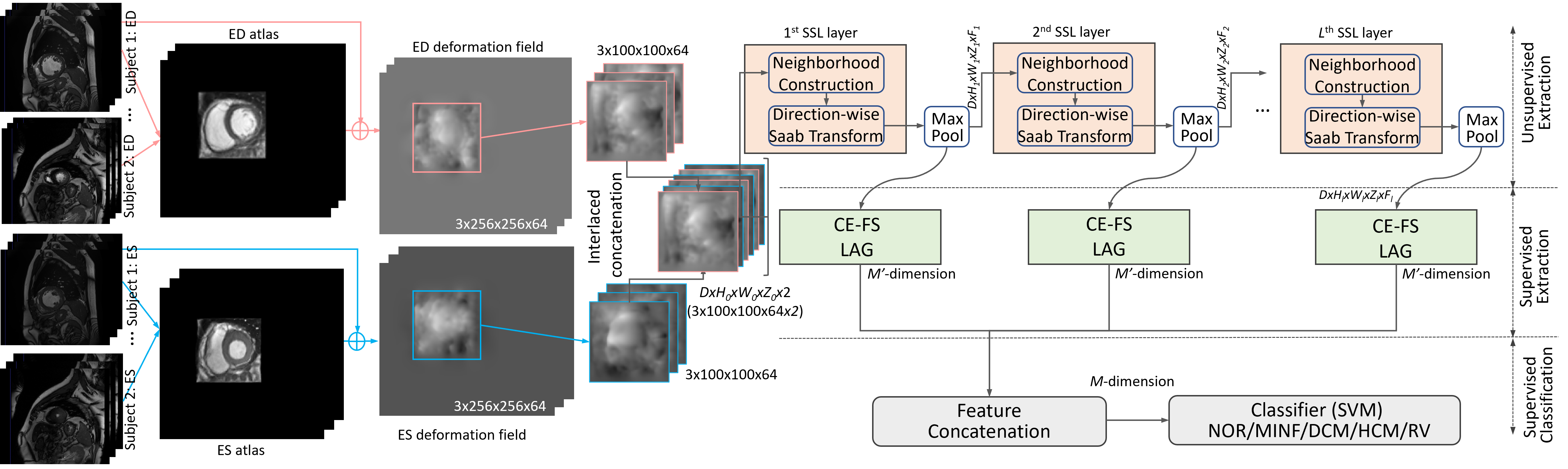} \vspace{-18pt}
  \caption{Illustration of our proposed SSL framework using $L$ stacked layers.}\label{fig1} \vspace{-3pt}
\end{figure*}


In this work, we propose to develop a lightweight and interpretable model based on successive subspace learning (SSL) for accurate and efficient CVD classification~\cite{rouhsedaghat2021successive}. We use two phases, including ED and ES, to represent the entire cardiac cycle. To encode objective and characteristic differences across individuals, we first construct a cardiac atlas from cine MRI using diffeomorphic groupwise registration. Second, we perform diffeomorphic registration between the atlas and individual subjects to yield 3D deformation fields in both phases. Since 3D deformation fields encode compression and expansion of each tissue point with respect to the mean representation, they can offer objective and detailed means of assessing CVD. Then, the interlaced concatenation of the deformation fields for both ED and ES phases is processed by our framework to explore the difference between the two phases. Our framework follows a multi-layer stacked design, in which each layer consists of the following modules: (i) direction-wise 3D neighborhood voxel construction for the sequential expansion of near-to-far neighborhood for local-to-global information extraction; (ii) unsupervised dimension reduction with direction-wise subspace approximation with adjusted bias (Saab) transform; (iii) supervised dimension reduction with novel class-wise entropy guided feature selection and label-assisted regression; and (iv) multi-layer feature concatenation for final classification with support vector machine (SVM). Notably, to address the sign confusion problem in DL, the Saab transform \cite{kuo2019interpretable} has been proposed to replace nonlinear activation units~\cite{kuo2016understanding}. More appallingly, its parameters are computed in a feedforward manner, which does not depend on the backpropagation and is thereby more mathematically interpretable~\cite{kuo2019interpretable,fan2020interpretability}.

The main contribution of this work is summarized as:

$\bullet$ To our knowledge, this is the first effort at exploring two-phase 3D deformation fields with an SSL framework for the CVD classification task. 

$\bullet$ Novel direction-wise Saab transform alongside class-wise entropy-guided feature selection is performed for efficient unsupervised/supervised dimension reduction for 4D deformation fields (3D spatial deformation+direction).

$\bullet$ Thorough evaluations on the ACDC2017 database show superior CVD classification performance of our framework over 3D CNN-based approaches with 140$\times$ fewer parameters.

\vspace{-5pt}
\section{Methodology}

\vspace{-5pt}
\subsection{Cardiac atlas and deformation field construction}
\vspace{-3pt}

Given ED and ES phases ($\mathbb{R}^{256\times256\times64}$) of cine MRI volumes from controls, we first construct a cardiac atlas for both ED and ES phases with group-wise diffeomorphic registration with cross-correlation as our similarity measure~\cite{avants2008symmetric}. Second, using cine MRI volumes in both ED and ES phases, we compute multi-phase 3D deformation fields to encode relative deformations with respect to the mean ED and ES atlas volumes. This is achieved by diffeomorphic registration between the ED and ES volumes of each individual and the ED and ES atlas volumes, respectively. As a result, we have the deformation fields ($\mathbb{R}^{3\times256\times256\times64}$) with three directions in each voxel. The region of interest (ROI) is cropped to $\mathbb{R}^{3\times100\times100\times64}$, and the interlaced concatenation ($\mathbb{R}^{3\times100\times100\times128}$) is used as input to our framework. 



\vspace{-5pt} 
\subsection{SSL based on two-phase deformation fields}
\vspace{-2pt}

Our framework follows a stacked design in a feedforward manner, which consists of $L$ cascade layers for local-to-global information extraction as shown in Fig.\ref{fig1}. Each layer has (i) direction-wise 3D neighborhood voxel construction, (ii) unsupervised Saab transform, (iii) supervised class-wise entropy-guided feature selection, and (iv) SVM with multi-layer features concatenation.

\vspace{-5pt}
\subsubsection{Unsupervised subspace approximation}
\vspace{-2pt}

First, the neighborhood voxel union is constructed to define the region to be explored neighboring spatial content. Instead of the conventional image-based SSL~\cite{chen2020pixelhop}, we propose a direction-wise 3D neighborhood voxel construction for our two-phase deformation fields. Each union has the size of $h\times w\times z$ and the step length of 1 in 3D space. Therefore, for an input sample with the size of $H\times W\times Z$, we have $(H-h+1)\times(W-w+1)\times(Z-z+1)$ neighborhood unions after incorporating a boundary effect. Of note, we have $H,W=100$, $Z=128$ in each direction of the first layer. Notably, the three directions are processed independently at the unsupervised feature extraction stage, which can be processed in parallel. Then, each union is flattened as a vector $x$ with the length of $h\times w \times z$, which is to be compacted to a $F_l$ dimension vector in the $l$-th layer with the Saab transform. 

The Saab transform can be seen as a variant of principal component analysis (PCA), and the transform resorts to the direct current (DC) and alternating current (AC) anchor vectors, by expressing $x$ with a compacted approximation feature $f$. Specifically, for the Saab transform in our $l$-th layer, we configure 1 DC vector and $F_l-1$ AC vectors. Of note, all of them have the length of $h\times w \times z$ as $x$. Therefore, we can formulate the $i$-th dimension of $f$ as an affine transform of $x$. Specifically, we have:\vspace{-5pt}
\begin{equation}
    f_i=a_i^Tx+b_i,~~~ i=0,1,\cdots, F_1-1,
\end{equation} 
where the bias term $b_i$ is a scalar~\cite{kuo2019interpretable}. Similar to PixelHop \cite{kuo2019interpretable}, we simply define $b_i\equiv d\sqrt{F_l}, d\in\mathbb{R}$, and split the anchor vectors into DC and AC vectors as:\vspace{-5pt}
\begin{equation}
\begin{aligned}
&\bullet~{\text{DC anchor vector}}~~a_0=\frac{1}{\sqrt{h\times w \times z}}(1,\cdots, 1)^T, \\
&\bullet~{\text{AC anchor vector}}~~a_i,~~i=1,\cdots, F_l-1. \label{eq:1}
\end{aligned}\end{equation}
We would expect the subspace of AC to be the orthogonal complement to the subspace of DC. Therefore, the AC component of $x$ can be formulated as $x_{AC}=x-x_{DC}$. Then, we apply PCA to $x_{AC}$, and choose the top $F_1-1$ principal components as our AC anchor vectors. An anchor vector operates on a region of the input data, and generates a scalar, which is similar to the convolution operation in CNNs. The anchor vector can be seen as a filter~\cite{kuo2019interpretable}. In addition, using several anchor vectors is similar to using numerous filters in contemporary DL. Instead of iteratively learned filters in DL with supervised backpropagation, the anchor vectors in our framework are unsupervised as those are defined using PCA.  

After the Saab transform of each union, we half the spacial size with the max-pooling operation and send it to the next SSL layer. With the multi-channel input, the neighborhood construction involves $F_1$ channels at each voxel position, which is processed in parallel simliar to PixelHop~\cite{kuo2019interpretable}. With the stacked SSL layers, the neighborhood unions are related to more original voxels to explore global features. Similarly, modern CNNs achieve a larger reception field, by using deeper layers. The detailed layer-wise unsupervised dimension reduction modules in our SSL model are provided in Table \ref{table:1}.

\vspace{-5pt}
\subsubsection{Supervised dimension reduction}
\vspace{-2pt}

The supervised dimension reduction module is used to (i) select the discriminative feature channels using the label supervision, and (ii) unify the shape of features in each layer for the subsequent concatenation. 
 
The $F_l$ channels in each voxel can have different importance for the CVD classification. Therefore, it can be helpful to apply a supervised feature selection. Following the processing of 2D images in PixelHop++~\cite{chen2020pixelhop++}, we propose to use direction-wise voxel cross-entropy guided feature selection (CE-FS). Specifically, we only keep the channels with small entropy of each class, which explicitly enforces the feature of a voxel in each channel to be similar~\cite{chen2020pixelhop++}. We use $p_k^c$ to denote the feature of the $k$-th voxel of a class in the $c$-th channel. Following~\cite{chen2020pixelhop++}, we can formulate the entropy of a sample as:\vspace{-5pt}
\begin{equation}
\begin{aligned}
\mathcal{H}=\sum_{n=1}^5 \mathcal{H}_n,~~\mathcal{H}_n= -\sum_{k} p_k^c {\text{log}} p_k^c, \label{eq:1}
\end{aligned}\end{equation}
where we index the involved class category with $n$, e.g., $n\in\{1,\cdots,5\}$. Then, the calculated entropy of all classes for each channel is descending ordered. Only the top 50\% channels with small entropy are kept for efficient classification. 
  
Then, the label-assisted regression (LAG)~\cite{chen2020pixelhop} is applied to each layer to extract the final layer-wise feature $v_l$ with a length of 25 dimensions consistently. 

\begin{table}[t]\vspace{-5pt}
\caption{The details of unsupervised modules in our 5-layer model for two-phase deformation fields CVD diagnosis} 
\centering 
\resizebox{1\columnwidth}{!}{%
\begin{tabular}{l | l | l} 
\hline  
Input Size&Type& Filter Shape   \\ [0.5ex] 
\hline 

$3\times[100\times100\times64\times2]$&D-Saab&  $3\times F_1$ kernels of $[3\times3\times6$]\\
$3\times[98\times98\times123\times F_1]$&MaxPool& (2$\times$2$\times$2)-(1$\times$1$\times$1)\\
\hline
$3\times[49\times49\times62\times F_1]$&D-Saab&  $3\times F_2$ kernels of $[3\times3\times3$]\\
$3\times[47\times47\times60\times F_2]$&MaxPool& (2$\times$2$\times$2)-(1$\times$1$\times$1)\\
\hline
$3\times[24\times24\times30\times F_2]$&D-Saab&  $3\times F_3$ kernels of $[3\times3\times3$]\\
$3\times[22\times22\times28\times F_3]$&MaxPool& (2$\times$2$\times$2)-(1$\times$1$\times$1)\\
\hline
$3\times[11\times11\times14\times F_3]$&D-Saab&  $3\times F_4$ kernels of $[3\times3\times3$]\\
$3\times[9\times9\times12\times F_4]$&MaxPool& (2$\times$2$\times$2)-(1$\times$1$\times$1)\\
\hline
$3\times[5\times5\times6\times F_4]$&D-Saab&  $3\times F_5$ kernels of $[3\times3\times3$]\\
$3\times[3\times3\times3\times F_5]$&MaxPool& (2$\times$2$\times$2)-(1$\times$1$\times$1)\\
\hline  
 
\end{tabular}
\label{table:1} 
}
\end{table}

\vspace{-5pt}
\subsubsection{Multi-level feature integration}
\vspace{-2pt}

With the extracted features $v_l$ in each layer with both unsupervised and supervised dimension reductions, we sequentially concatenate them in all three directions as our final sample-wise CVD feature. We empirically choose SVM as our CVD classification model, which has been well established and has been widely used for classification purposes. Of note, as a supervised machine learning model, SVM is trained with the training datasets. 

\vspace{-5pt}
\section{Experiments}
\vspace{-2pt}

\begin{figure*}[t!]
\centering 
\includegraphics[width=18cm]{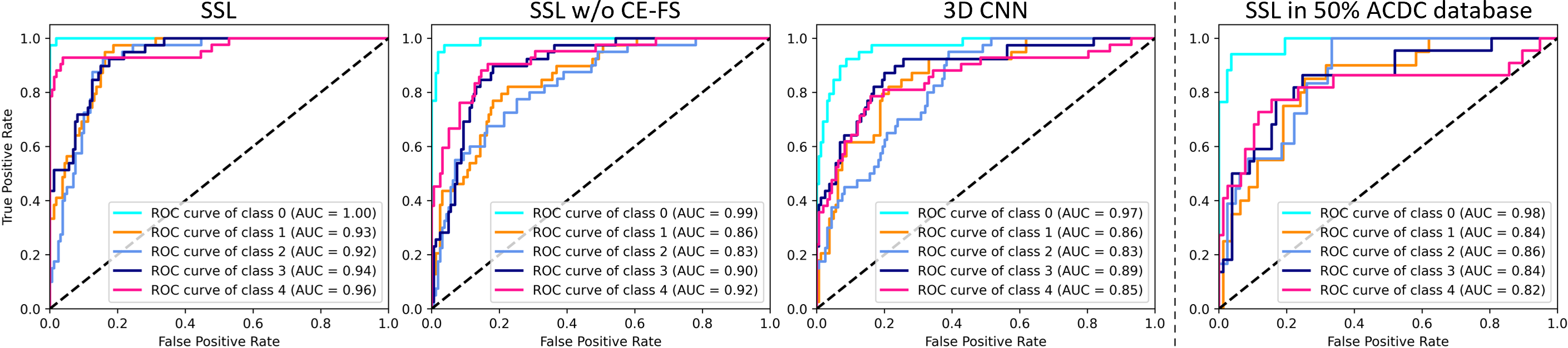}  \vspace{-15pt}
\caption{Comparison of the ROC curves of (a) SSL, (b) SSL w/o CE-FS, (c) 3D CNN, and (d) SSL with half ACDC2017 data. Classes 0 to 4 represent the NOR, MINF, DCM, HCM, and RV, respectively.}\label{figexp1} 
\end{figure*} 

We evaluate our framework on the ACDC2017 database, which consists of a total of 99 subjects, including 20 normal subjects (NOR), 20 myocardial infarction (MINF) subjects, 19 dilated cardiomyopathy (DCM) subjects, 20 hypertrophic cardiomyopathy (HCM) subjects, and 20 abnormal right ventricle (RV) subjects. For each subject, cine MRI short-axis slices, including ED and ES phases, were acquired. Notably, we consistently resampled the training samples with the size of $256\times256\times64$. We followed the 5-fold cross-validation to split the subjects with uniform categorical sampling. 
 

We empirically used five SSL layers and configured $F_1$=5, $F_2=$5, $F_3=$15, $F_4=$20, and $F_5=$25. Notably, the maintained energy ratio in the unsupervised dimension reduction module was controlled by the quantity of the Saab filters, i.e., $F_l$. In addition, we implemented the compared DL models using the Pytorch toolkit. 

As comparison methods, we used the ResNet-based 3D CNN~\cite{korolev2017residual} and 3D self-attention~\cite{wang2021automated} based classification model for the 3D CVD classification task. Specifically, the ResNet \cite{he2016identity} and self-attention network~\cite{wang2021automated} have been widely used for 3D medical classification tasks~\cite{korolev2017residual,singh20203d,wang2021automated}. To adapt these methods for our multi-direction 3D deformation fields, we configured independent convolutional layers for each direction as~\cite{nie2019multi}. Then, the extracted direction-wise features in the first fully connected (FC) layer were concatenated. We used the threshold of 0.5 for the final prediction of the DL approaches. Notably, we fixed 3 FC layers and validated the convolutional layers of 3D CNN and self-attention layers from three to ten layers. We chose the best-performed layer for ResNet-based 3D CNN and 3D self-attention, i.e., five or six layers, respectively.


 \begin{table}[t!]\vspace{-10pt}
\caption{Comparison of the 5-fold cross-validation accuracy on different proportion of ACDC2017 database}\vspace{+2pt}
\centering 
\resizebox{1\columnwidth}{!}{%
\begin{tabular}{l | c | c   c   c } 
\hline 
Methods& \textbf{Parameters}  & ACDC & 75\%ACDC & 50\%ACDC  \\  
\hline 

ResNet 3D CNN& $\sim$5.6 M & 0.88 & 0.83 & 0.68\\
3D Self Attention& $\sim$8.5 M & 0.85 & 0.78& 0.72 \\\hline
SSL w/o CE-FS&$\sim$\textbf{0.04 M} & 0.92& 0.90 & 0.84  \\
SSL w/o IC&$\sim$\textbf{0.04 M} & 0.91  & 0.87 & 0.84 \\\hline
SSL& $\sim$\textbf{0.04 M}& \textbf{0.95} & \textbf{0.91} & \textbf{0.86}  \\\hline
\end{tabular}
}\label{table:2}
\end{table}

In Fig.2, we plot the receiver operating characteristic (ROC) curves of each CVD class. We can see that our framework outperformed the compared 3D CNN models. As well, the area under the ROC curves of each class are also reported in Fig. 2. To facilitate the following LAG modules, CE-SF is applied as a preliminary supervised dimension reduction method. We provided the ablation study of CE-SF (i.e., SSL w/o CE-FS) with doubled feature vectors for the subsequent LAG model, which yielded inferior performance with an increase of 18\% overall processing time. 
 
In Table 2, we show the accuracy of different methods. To further investigate the performance with a smaller number of datasets, we uniformly selected 75\% and 50\% of 100 subjects for 5-fold validation. Our SSL outperformed the compared DL-based models consistently by a larger margin especially in the case of a smaller number of datasets. Furthermore, we provided the ROC curves using 50\% of the ACDC2017 database. We can see that our SSL model can be robust even with very limited training data. Of note, the number of parameters of our SSL framework was about 140$\times$ fewer than the compared 3D CNN architectures. The much fewer parameters can alleviate the difficulty of using a limited number of training datasets. Experimental results from our ablation study of SSL w/o CE-FS also showed its efficacy. Moreover, we provided the ablation study of the interlaced concatenation, which is denoted as SSL w/o IC. The performance drop compared with the SSL demonstrates that exploring the difference between ED and ES phases in an early voxel union was necessary.

\begin{table}[t!]\vspace{-10pt}
\caption{Sensitivity study of the stacked layers. The 5-fold cross-validation accuracy on full ACDC2017 is reported}  \vspace{+2pt}
\centering 
\resizebox{0.9\columnwidth}{!}{%
\begin{tabular}{l | c | c | c  | c | c | c  } 
\hline  
Layers & 3 & 4 & 5 & 6 & 7 & 8 \\  
\hline 

ResNet 3D CNN & 0.76 & 0.86 & 0.88  & 0.87 & 0.85 & 0.83 \\
3D Self Attention &0.79&  0.82 & 0.85 & 0.85  & 0.83 & 0.80\\\hline
SSL&0.85& 0.92 & 0.95 &  0.95 & 0.95 & 0.96 \\\hline
\end{tabular}
}\label{table:3}
\end{table}
 
The number of SSL layers is important for our CVD classification task to balance performance and efficiency. In Table 3, we show the detailed sensitivity study using a different number of SSL layers. The performance was relatively stable after four layers. Notably, there was a performance drop for deeper DL methods with more to-be-trained parameters.

\vspace{-2pt}
\section{Conclusion}
\vspace{-2pt}

In this work, we proposed a novel lightweight and interpretable framework with two-phase deformation fields for the CVD classification task. In particular, the class-wise entropy-guided feature selection was proposed to achieve accurate classification. We evaluated our framework on the ACDC2017 database using a different number of training datasets, demonstrating superior performance over 3D CNN and self-attention DL models, with about 140$\times$ fewer parameters. It is important to note that our framwork is based on the feedforward Saab transform for which our framework is deemed more interpretable than CNNs, which rely on backpropagation. Taken together, our framework offers the potential to be used for clinical practice with a limited number of imaging data. 
 
\vspace{-2pt}
\section{Acknowledgment} The authors would like to thank Dr. Reza Nezafat for his valuable
insights and helpful discussions.
\vspace{-2pt}

\section{COMPLIANCE WITH ETHICAL STANDARDS}  
This research study was conducted retrospectively using human subject data made available in open access by
\href{https://acdc.creatis.insa-lyon.fr/description/}{Automated Cardiac Diagnosis Challenge (ACDC)'17}.

\bibliographystyle{IEEEbib}
\bibliography{refs}

\end{document}